\documentclass[useAMS,usenatbib]{mn2e}

\usepackage{amssymb,natbib,graphicx,subfigure,color}

\newcommand{\be}{\begin{equation}}
\newcommand{\ee}{\end{equation}}
\newcommand{\bea}{\begin{eqnarray}}
\newcommand{\eea}{\end{eqnarray}}
\newcommand{\equ}[1]{eq.~(\ref{e:#1})}
\newcommand{\equs}[1]{eqs.~(\ref{e:#1})}
\newcommand{\Equ}[1]{Eq.~(\ref{e:#1})}
\newcommand{\Equs}[1]{Eqs.~(\ref{e:#1})}

\newcommand{\ifm}[1]{\relax\ifmmode#1\else$\mathsurround=0pt #1$\fi}

\newcommand{\sggg}[1]{\textcolor{green}{[]}}

\def\LCDM{$\Lambda\mbox{CDM}$}

\def\tsf{t_{\rm SF}}

\def\pc{{\rm\thinspace pc}}
\def\kpc{{\rm\thinspace kpc}}

\def\kms{{\rm\thinspace km s}^{-1}}

\def\erg{{\rm\thinspace erg}}

\def\g{{\rm\thinspace g}}

\def\Msun{\hbox{$\rm\thinspace M_{\odot}$}}
       
\def\yr{{\rm\thinspace yr}}
\def\Myr{{\rm\thinspace Myr}}
\def\Gyr{{\rm\thinspace Gyr}}
\def\Msunyr{\Msun\yr^{-1}}
\def\Msunpc2{{\Msun\pc}^{-2}}
\def\Msunyrkpc2{{\Msun\yr^{-1}\kpc}^{-2}}
\def\magarcsec2{{\rm\thinspace mag\thinspace arcsec}^{-2}}

\newcommand{\apj}{ApJ}
\newcommand{\apjl}{ApJL}

\newcommand{\aj}{AJ}
\newcommand{\mnras}{MNRAS}
\newcommand{\aap}{A\&A}

\newcommand{\araa}{ARA\&A}
\newcommand{\nat}{Nature}

\newcommand{\ltsima}{$\; \buildrel < \over \sim \;$}
\newcommand{\lsim}{\lower.5ex\hbox{\ltsima}}
\newcommand{\gtsima}{$\; \buildrel > \over \sim \;$}
\newcommand{\gsim}{\lower.5ex\hbox{\gtsima}}
\newcommand{\prop}{\propto}

\def\la{\langle}

\def\Vrot{V_{\rm rot}}
\def\tdyn{t_{\rm dyn}}
\def\tdis{t_{\rm dis}}
\def\tsf{t_{\rm SF}}
\def\gamdis{\gamma_{\rm dis}}
\def\gamdisII{\gamma_{\rm dis,2}}
\def\etaw{\eta}
\def\gaminf{\gamma_{\rm inf}}
\def\gamrcl{\gamma_{\rm rcl}}
\def\epsSFR{\epsilon_{\rm SF}}

\def\epsSFRII{\epsilon_{\rm SF,0.02}}

\begin{document}

\title[Cosmological inflow and disc turbulence]{On the effect of cosmological inflow on turbulence and instability in galactic discs}

\author[Genel, Dekel \& Cacciato]{Shy Genel$^{1,2}$\thanks{E-mail: sgenel@cfa.harvard.edu}, Avishai Dekel$^{3}$, Marcello Cacciato$^{3}$\\
$^{1}$Harvard-Smithsonian Center for Astrophysics, 60 Garden Street, Cambridge, MA 02138, USA\\
$^{2}$School of Physics and Astronomy, Tel Aviv University, Tel Aviv 69978, Israel\\
$^{3}$Racah Institute of Physics, The Hebrew University, Jerusalem 91904, Israel}

\maketitle

\label{firstpage}

\begin{abstract}
We analyse the evolution of turbulence and gravitational instability of a galactic disc in a quasi-steady state governed by cosmological inflow. We focus on the possibility that the coupling between the in-streaming gas and the disc is maximal, e.g., via dense clumps, and ask whether the streams could be the driver of turbulence in an unstable disc with a Toomre parameter $Q \sim 1$. Our fiducial model assumes an efficiency of $\sim 0.5$ per dynamical time for the decay of turbulence energy, and $\sim 0.02$ for each of the processes that deplete the disc gas, i.e., star formation, outflow, and inflow within the disc into a central bulge. In this case, the in-streaming drives a ratio of turbulent to rotation velocity $\sigma/V \sim 0.2-0.3$, which at $z \sim 2$ induces an instability with $Q \sim 1$, both as observed. However, in conflict with observations, this model predicts that $\sigma/V$ remains constant with time, independent of the cosmological accretion rate, because mass and turbulence have the same external source. Such strongly coupled cosmological inflow tends to stabilize the disc at low $z$, with $Q\sim$ a few, which may be consistent with observations. The instability could instead be maintained for longer, with a properly declining $\sigma/V$, if it is self-regulated to oscillations about $Q\approx1$ by a duty cycle for disc depletion. However, the `off' phases of this duty cycle become long at low $z$, which may be hard to reconcile with observations. Alternatively, the coupling between the in-streaming gas and the disc may weaken in time, reflecting an evolving nature of the accretion. If, instead, that coupling is weak at all times, the likely energy source for self-regulated stirring up of the turbulence is the inflow within the disc down the potential gradient (studied in a companion paper).
\end{abstract}


\section{Introduction}
\label{s:intro}
The basic kinematical properties of galaxies are the average rotation 
velocity $\Vrot$, and the velocity dispersions $\sigma$ of their components, namely
the random motions of stars and the gas turbulence in the interstellar medium (ISM).
Disc galaxies are supported against gravity by rotation. 
At low redshift, massive discs have gas velocity dispersions of $\sim 10\kms$
\citep{DibS_06a}, with $\sigma/\Vrot\approx0.05-0.1$. Their
stellar velocity dispersion is typically at the level of tens of $\kms$
and it varies with stellar surface density
\citep{BottemaR_93a,vanderKruitP_10a,WestfallK_11a}.
Observations show larger
gas velocity dispersions at higher redshifts \citep{EpinatB_10a,DaviesR_11a},
such that typical massive disc galaxies at $z\approx2$ have
$\sigma\approx30-80\kms$ and $\sigma/\Vrot\approx0.15-0.3$
\citep{ErbD_04a,FoersterSchreiberN_06a,CresciG_09a}.

Galactic discs are often assumed to maintain marginal gravitational instability, with
a Toomre $Q$ parameter $Q \sim 1$ \citep{ToomreA_64a}. 
In this case, 
\be
\frac{\sigma}{\Vrot} \approx (2\nu)^{-1/2} \delta \, ,
\label{e:delta}
\ee
where $\delta$ is the mass fraction in cold disc within the disc radius,
\be
\delta \equiv \frac{M_{\rm disc}}{M_{\rm tot}} \, ,
\label{e:delta_def}
\ee
and $\nu$ is a factor of order unity that depends on the shape of the
rotation curve,
\be 
\nu \equiv 1+\frac{\log \Vrot}{\log r} \simeq 1 
\label{e:nu}
\ee
\citep{DekelA_09b}.
\Equ{delta} explains why at high redshift the high fraction of cold disc, 
in terms of gas and young stars, requires a high value of $\sigma/\Vrot$ 
in order to maintain $Q \sim 1$.
The instability of discs with high $\delta$ and $\sigma/\Vrot$ is more 
`violent', in the sense that the structures are larger and the dynamical
processes are faster 
\citep{GenzelR_08a,DekelA_09b,GenzelR_11a,CeverinoD_11a,BournaudF_11a,BournaudF_12a}.
The perturbations associated with the instability, 
in the form of extended transient features and bound clumps,  
reflect the larger characteristic Toomre mass, 
\be
\frac{M_{\rm clump}}{M_{\rm disc}} \simeq \frac{1.2}{\nu}\,
\left(\frac{\sigma}{\Vrot} \right)^2 \, .
\label{e:Mclump}
\ee
Torques between the perturbations drive angular momentum out and
generates mass inflow, partly as clump migration and partly as inflow of
inter-clump mass. The timescale for inflow can be estimated in several 
different ways \citep{GammieC_01a,DekelA_09b} to be  
\be
\frac{t_{\rm inf}}{\tdyn} \simeq 5\,\left( \frac{\Vrot}{\sigma}\right)^2 \, ,
\label{e:tmig}
\ee
where
\be
\tdyn \equiv \Omega^{-1} = \frac{R}{\Vrot} \, ,
\label{e:tdyn}
\ee
$\Omega$ is the angular velocity and $R$ is the effective radius of the disc.
Thus, a higher $\sigma/\Vrot$ is associated with a faster inflow in the disc.

Since ISM turbulence decays on one or a few dynamical timescales
\citep{MacLowM_98a,StoneJ_98a,GammieC_01a,ElmegreenB_04a},
there must exist a continuous energy source that maintains the turbulence over cosmic time.
However, the nature of this energy source is highly debatable.
The mechanisms that drive the ISM turbulence could be divided into three
kinds. First, {\it stellar feedback}, such as
supernova feedback and radiative feedback from stars,
which deposit energy and momentum into the ISM.
Second, the energy source could be the gravitational energy released 
by the instability-driven inflow down the potential gradient within the disc, 
which is a natural mechanism for self-regulating the instability at $Q \sim 1$
\citep{WadaK_02a,AgertzO_09b,BournaudF_10a,KrumholzM_10b,CacciatoM_11a}.
Third, the driver of turbulence could be the kinetic energy transferred from
the cosmological inflow of clumpy gas.
At high redshift, the gas streams in as supersonic streams that follow the
filaments of the cosmic web. The streams,
consisting of merging galaxies and a smoother component,
penetrate through the halo to the vicinity of the central disc
where they deposit a fraction of their energy and momentum
\citep{BirnboimY_03a,KeresD_05a,DekelA_06a,KeresD_09a,DekelA_09a}.
At low redshift, the flow consists of cold gas clouds that `rain' 
from the hot halo onto the disc \citep{MallerA_04a,DekelA_08a,KeresD_09b}.
Such flows may be efficiently converted into
turbulence \citep{KlessenR_10a}, provided that the in-streaming gas has comparable density
to the disc ISM (e.g.~\citealp{DekelA_09b}).

The role played by stellar feedback in driving the ISM turbulence is
controversial.
On one hand, theoretical estimates and numerical simulations argue that 
stellar feedback could be the main driver of turbulence in local massive
galaxies
\citep{MacLowM_04a,KimJ_01a,DibS_06a,deAvillezM_07a} as well as in high-redshift
galaxies \citep{HopkinsP_11a}.
On the other hand, other estimates and simulations argue that stellar 
feedback is unlikely to drive a velocity dispersion larger than $\approx10\kms$
\citep{JoungM_09b,DekelA_09b,BournaudF_10a,OstrikerE_11a}.

Unlike stellar feedback, where the energy emerges from nuclear processes inside
stars, both the second and third mechanisms refer to `gravitational heating',
where gravitational potential energy is released as a result of infall
into a potential well \citep{DekelA_08a,KhochfarS_08a}.
The second mechanism, based on the inflow within the disc,
is determined by the self-regulated disc
instability, and is the topic of a companion paper by \citet{CacciatoM_11a}.
The third, based on mass streaming in from outside the disc, represents
an external source of energy that is determined by the cosmic growth of
structure and is independent of the disc instability.
In this work we focus on this external mechanism.

So far, different authors seem to have reached different conclusions 
concerning the possible role played by these `gravitational heating' mechanisms at $z \sim 2$
\citep{GenzelR_08a,KhochfarS_09a,LehnertM_09a,ElmegreenB_10a,KlessenR_10a,KrumholzM_10b}.
This uncertainty naturally stems from the fact that  
the power provided by cosmological in-streaming is 
in the same ball park as the turbulence dissipation rate.
This motivates the more detailed analysis presented in this paper.

The evolution of the gas mass in a disc galaxy can be described 
as a quasi-steady-state solution of a simple differential equation
of mass conservation
\citep[][see below]{FinlatorK_08a,DekelA_09b,BoucheN_10a,DuttonA_10a,DaveR_11b,DaveR_11a}. 
In turn, the generation of turbulence and its dissipation in a steady state
is governed by an analogous equation of energy conservation
(e.g.~\citealp{KhochfarS_09a,ElmegreenB_10a,KlessenR_10a}).
Together, these equations help constrain the disc instability, 
as the $Q$ parameter depends on disc mass and turbulence.

There is numerical and observational evidence for marginal instability with
$Q \sim 1$ in disc galaxies.  
Simulations reveal $Q\approx1$ in high-redshift, gas-rich discs
\citep{ImmeliA_04b,CeverinoD_09a,GenelS_11a,HopkinsP_11a}, and
$Q\approx2-3$ in low-redshift stellar-dominated discs 
\citep{HohlF_71a,AthanassoulaE_86a,BottemaR_03a,MartigM_09a}.
Similar estimates are obtained from observed discs, both in local galaxies
\citep{LeroyA_08a,WestfallK_11a,vanderKruitP_11a,YimK_11a,WatsonL_12a} and in
$z\sim 2$ discs \citep{GenzelR_11a}.
However, the parameter $Q$ applies to linear perturbation theory and 
caution is required in comparing it with estimates from the high non-linearly evolved
(observed and simulated) galaxy discs.

The instability of a disc and the associated level of turbulence may be 
coupled via a self-regulation mechanism \citep[e.g.,][]{DekelA_09b}. 
If the turbulence is driven by internal processes that themselves depend on the
disc instability, such as star formation or internal torques that cause 
mass inflow, the system may relax into a steady-state by a self-regulation loop.
In this case the disc maintains marginal instability, $Q\sim1$, as the
turbulence adjusts itself to the proper value dictated by the gas surface
density and the angular velocity. 
In a companion paper (\citealp{CacciatoM_11a}, see also \citealp{KrumholzM_10b}),
we impose $Q=1$, and analyse the steady state
solution of the mass and energy equations under the assumption that the 
energy source for driving the turbulence is the inflow down the potential 
gradient within the disc. The rate of this inflow adjusts itself to compensate 
for the turbulence dissipative losses such that $Q=1$ is maintained.
In \citet{CacciatoM_11a} we address the instability of a two-component disc,
with gas and stars of different velocity dispersions that gradually exchange
mass. We find there that discs tend to `stabilize' at low redshift as 
the disc becomes dominated by the `hot' stellar component.
\citet{ForbesJ_12a} study a similar scenario including radial variations 
within the disc.

In this paper, we study the steady-state solution of similar mass and energy
conservation equations, but focus on an external energy source, carried by the
cosmological in-streaming of gas. We address two main cases.
First, a case where the system is governed by the external source alone
and the instability is not self-regulated, and where the efficiencies of the
various physical processes are constant in time. 
Second, an alternative case where the instability is
self-regulated by a duty-cycle for instability and star formation. 

This paper is organized as follows. 
In Section \ref{s:conservation_eqs} we present the equations of mass and energy
conservation and their steady-state equations, and introduce our
parameterization of the relevant physical scenarios.
In Section \ref{s:caseI} we investigate the non-self-regulated 
case I, with fixed efficiencies of the physical processes, and predict
the evolution of $Q$ and $\sigma/\Vrot$.
In Section \ref{s:solutions_Q_1} we study the self-regulated case II,
where we impose $Q=1$ and introduce a duty cycle for instability and star
formation.
In Section \ref{s:literature} we put our results in the context of other
results from the literature.
In Section \ref{s:summary} we conclude and discuss our results.

\section{Equations for mass and energy conservation}
\label{s:conservation_eqs}
In this section we present the basic equations for conservation of mass and turbulent energy in a gaseous galactic disc. Source and drain terms for gas mass and gas turbulent energy are identified. As a result, the gas mass and gas velocity dispersion that characterizes the turbulence are computed in a steady-state solution as a function of the incoming supply rate of cosmological gas and the parameters that characterize the various relevant physical processes.

\subsection{The backbone steady state model}
\label{s:backbone}
The basic equation for the gas mass budget of a galactic disc is
\be
\dot{M}_{\rm g}=\dot{M}_{\rm cosmo}-\dot{M}_{\rm sink},
\label{e:mass_conservation}
\ee
where $M_{\rm g}$ is the disc gas mass, $\dot{M}_{\rm cosmo}$ is the external source term that represents the cosmological gas inflow rate, and $\dot{M}_{\rm sink}$ is the sum of different kinds of `sinks' that empty the disc of its gas, including star formation, galactic outflows, and inflows inside the disc into the bulge. \Equ{mass_conservation} has a very simple and instructive solution if the sink terms can be written as $\dot{M}_{\rm sink}=M_{\rm g}\tau^{-1}$, i.e.~if they are proportional to the gas mass itself with a `sink timescale' proportionality factor\footnote{As we discuss later, $\tau$ is related to the dynamical time of the disc.}. If $\dot{M}_{\rm cosmo}$ and $\tau$ vary on a timescale longer than $\tau$, the solution is
\bea
M_{\rm g}&=&\dot{M}_{\rm cosmo}\tau(1-e^{-t/\tau})
\label{e:mass_conservation_solution_Mgas}\\
\dot{M}_{\rm g}&=&\dot{M}_{\rm cosmo}e^{-t/\tau}.
\label{e:mass_conservation_solution}
\eea
For $t\gg\tau$, it reduces to a steady state solution with
\bea
\dot{M}_{\rm sink}&\approx&\dot{M}_{\rm cosmo}
\label{e:mass_conservation_asymptotic_solution_Msink}\\
\dot{M}_{\rm g}&\approx&0
\label{e:mass_conservation_asymptotic_solution}\\
M_{\rm g}&\approx&\dot{M}_{\rm cosmo}\tau \, ,
\label{e:mass_conservation_asymptotic_solution_Mg}
\eea
in which the sink term $\dot{M}_{\rm sink}$ adjusts itself to match the external source term $\dot{M}_{\rm cosmo}$ (see \citealp{BoucheN_10a}). The range of validity of this solution is discussed in detail in Appendix \ref{s:quasi_steady_state_validity}.

Assuming that the disc has reached the steady state, we use the results from mass and energy conservation to derive the turbulent velocity as follows. We start by considering, for simplicity, only the star formation (SF) part in the gas mass sink term, in the form
\be
\dot{M}_{\rm SF}=\frac{M_{\rm g}}{\tsf} \, ,
\label{e:backbone_SFR_law}
\ee
and obtain from mass conservation
\be
\dot{M}_{\rm SF}=\dot{M}_{\rm cosmo} \, ,
\label{e:backbone_SFR_equilibrium}
\ee
and
\be
M_{\rm g}=\dot{M}_{\rm cosmo}\tsf.
\label{e:backbone_Mg}
\ee

In analogy, for the turbulent energy $E_{\rm turb}$ we consider a sink term in the form of a dissipation rate,
\be
\dot{E}_{\rm dis}=\frac{E_{\rm turb}}{\tdis}.
\label{e:backbone_dissipation_law}
\ee
This leads in steady state, in analogy to \equ{backbone_SFR_equilibrium}, to
\be
\dot{E}_{\rm dis}=\dot{E}_{\rm cosmo} \, ,
\label{e:backbone_dissipation_equilibrium}
\ee
where $\dot{E}_{\rm cosmo}$ is the rate of in-streaming energy. The analogue to \equ{backbone_Mg} is then
\be
E_{\rm turb}=\dot{E}_{\rm cosmo}\tdis.
\label{e:backbone_Eturb}
\ee
We approximate $E_{\rm turb}\approx M_{\rm g}\sigma^2$ and
$\dot{E}_{\rm cosmo}\approx\dot{M}_{\rm cosmo}V_{\rm in}^2$,
where the in-streaming velocity $V_{\rm in}$, as well as the rotational velocity $\Vrot$, are assumed to be comparable to the virial velocity of the halo \citep{DekelA_09a}, and the conversion of
in-streaming kinetic energy to disc turbulence is assumed to be efficient. \Equs{backbone_Mg} and (\ref{e:backbone_Eturb}) then yield
\be
\frac{\sigma}{\Vrot}=\sqrt{\frac{\tdis}{\tsf}} \, .
\label{e:backbone_V_over_sigma}
\ee

A very interesting feature of \equ{backbone_V_over_sigma} is that $\sigma/\Vrot$ turns out to be {\it independent of the cosmological accretion rate itself}. This unique feature stems from the facts that (a) in steady state both the sinks of SFR and dissipation rate adjust themselves to the corresponding supply rates, (b) in our current model, the cosmological supply is a common source for both the mass and turbulent energy of the disc, such that the cosmological input always provides the same {\it specific} turbulent energy, and (c) this supply is determined externally, independent of the conditions in the disc. The implication of this special property of our current model is that $\sigma/\Vrot$ is expected to be invariant under variations in the mass input rate, which is probably the main source of variation in the galaxy properties related to disc instability, both between different galaxies and as a function of time in the history of each individual galaxy.

Both timescales in \equ{backbone_V_over_sigma} are expected to be related to the dynamical time of the disc,
\be
\tdis\equiv\gamdis\tdyn \, ,
\label{e:gamdis}
\ee
with $\gamdis$ a constant parameter with a likely value in the range $1-3$ \citep{MacLowM_98a,GammieC_01a}, and
\be
\tsf\equiv \frac{\tdyn}{\epsSFR} \, ,
\label{e:epsilon}
\ee
with $\epsSFR\approx0.02$ \citep{SilkJ_97a,GenzelR_10a}. With these fiducial values \equ{backbone_V_over_sigma} becomes
\be
\frac{\sigma}{\Vrot}=\sqrt{\epsSFR\gamdis}\approx 0.2 \, .
\label{e:backbone_V_over_sigma_with_value}
\ee
In addition to the uncertainty in these parameters, several numerical factors of order unity have been omitted in this simple derivation, which will be recovered in Section \ref{s:energy_steady_state}. Despite the fact that the numerical values of $\epsSFR$, $\gamdis$ and the other parameters are not known to great accuracy, we learn from \equ{backbone_V_over_sigma_with_value} that the available power in the external accretion is in the same ballpark as the power required at $z \sim 2$ for maintaining the turbulence in the discs. As long as the conversion efficiency of that energy into turbulent energy is high, this could in principle be the main driver of disc turbulence. At low redshift, the cosmological accretion carries more than enough energy to maintain the observed $\sigma/\Vrot\approx0.05$.

As an aside, the contribution of stellar feedback to driving turbulence in the disc can be estimated in a similar way, replacing the gravitational potential $V_{\rm in}^2$ by the energy provided by stars per unit stellar mass formed, $V_{\rm FB}^2$. Thus $\dot{E}_{\rm cosmo}\approx\dot{M}_{\rm cosmo}\Vrot^2$ is replaced by $\dot{E}_{\rm FB}=\dot{M}_{\rm SF}\cdot V_{\rm FB}^2$. With $\sim 10^{51}\erg$ released by each supernova and one supernova per $100\Msun$ of stars formed, the released energy corresponds to $V_{\rm FB} \sim 700\kms$, well above what is needed for driving the observed turbulence. However, the vast majority of the energy emitted is radiated away in the initial phases of the supernova evolution \citep{DekelA_86a,ThorntonK_98a}, such that the energy available to be deposited in the ISM corresponds to only $V_{\rm FB} \sim 100\kms$, lower than the required energy. Furthermore, a large fraction of that energy is likely to drive outflows from the disc rather than turbulence inside the disc \citep{MacLowM_99a,JoungM_09b,OstrikerE_11a}.

\subsection{Mass steady state}
\label{s:mass_steady_state}
We now generalize the mass sink term in \equ{mass_conservation} to include several additional processes, and have at steady state
\be
\dot{M}_{\rm g}=\dot{M}_{\rm cosmo}-\dot{M}_{\rm SF}-\dot{M}_{\rm w,out}+\dot{M}_{\rm w,in}-\dot{M}_{\rm inf}=0.
\label{e:mass_steady_state}
\ee
In the following, we describe the parameterization of the various sink terms. Table \ref{t:parameters} gives an overview of all model parameters.
\begin{itemize}
\item
The star-formation rate (SFR) is defined as
\be
\dot{M}_{\rm SF}\equiv D\epsSFR\frac{M_{\rm g}}{\tdyn},
\label{e:SFR_def}
\ee
where $D$ is a duty cycle (with a fiducial value of $1$, to be discussed in Section \ref{s:D}) and $\epsSFR$ is the fraction of the gas that turns into stars every dynamical time $\tdyn$ whenever star formation is `on' ($D=1$), as in \equ{epsilon}.
\item
It is assumed that stellar feedback blows galactic winds at a rate that is proportional to the SFR with a mass-loading factor $\etaw$, namely
\be
\dot{M}_{\rm w,out}\equiv\etaw\cdot\dot{M}_{\rm SF},
\label{e:Mout}
\ee
where typically $\etaw\sim1$. Galactic winds can also be driven by feedback from an Active Galactic Nucleus (AGN; e.g.~\citealp{NesvadbaN_08a}), which, for simplicity, we include in the parameter $\etaw$, since AGN activity and star formation are often concurrent.
\item
It is assumed that some fraction $\gamrcl$ of the mass that was blown out into the wind comes back as an instantaneous fountain:
\be
\dot{M}_{\rm w,in}\equiv \gamrcl \cdot \etaw \cdot \dot{M}_{\rm SF} \, .
\label{e:Mwindin}
\ee
The fiducial value we choose is $\gamrcl=0$, namely no returning winds.
\item
Disc instability is associated with torques that drive angular momentum out and mass in \citep{GammieC_01a}, to form a bulge at the disc centre. Part of this is clump migration, whose rate could be computed by dynamical friction or clump-clump interactions. The inflow rate can be expressed as
\be
\dot{M}_{\rm inf}\equiv D\gaminf \frac{M_{\rm g}}{\tdyn}=\gaminf\epsSFR^{-1}\dot{M}_{\rm SF} \, ,
\label{e:Mmig_def}
\ee
where $\gaminf$ is the inflow `efficiency' per dynamical time \citep[][eqs. (19) and (24)]{DekelA_09b}. Based on \equ{tmig}, we estimate $\gaminf\approx0.02$ at $z\sim2$, and $\gaminf\approx0.001$ at $z\sim0$. To avoid underestimating the effect of disc inflows, we use a fiducial value of $\gaminf\epsSFR^{-1}=1$. The parameter $D$ in \equ{Mmig_def} is the same duty cycle as in \equ{SFR_def} for the SFR, assumed to be determined by the instability duty cycle.
\end{itemize}

By solving \equ{mass_steady_state}, we obtain the steady-state solution
\be
M_{\rm g}=\frac{\dot{M}_{\rm cosmo}\tdyn}{D\epsSFR[1+\etaw(1-\gamrcl)+\gaminf\epsSFR^{-1}]}.
\label{e:Mg}
\ee
The contribution of each term is easy to understand qualitatively. The gas mass is proportional to the cosmological supply rate of gas. More vigorous outflows (large $\etaw$) reduce the gas mass in the disc, unless they are largely recycled (large $\gamrcl$). A duty cycle with longer `off' phases (small $D$) leaves more gas in the disc. A higher efficiency of star formation (large $\epsSFR$) reduces the gas mass. A stronger inflow inside the disc (large $\gaminf$) also reduces the disc gas mass. Using \equ{SFR_def}, we also obtain 
\be
\dot{M}_{\rm SF}=\frac{\dot{M}_{\rm cosmo}}{1+\etaw(1-\gamrcl)+\gaminf\epsSFR^{-1}}.
\label{e:SFR}
\ee

\begin{table*}
\caption{Model parameters.}
\label{t:parameters}
\begin{tabular}{cclc}
\hline\hline 
Parameter  & Fiducial Value & Definition & Equation \\
\hline
$\epsSFR$  & 0.02 & Star-formation efficiency per disc dynamical time & (\ref{e:SFR_def}) \\
$D$        & 1    & Duty cycle for instability, star-formation and inflows & (\ref{e:SFR_def}),(\ref{e:Mmig_def}) \\
$\etaw$    & 1    & Wind mass-loading factor & (\ref{e:Mout}) \\
$\gamrcl$  & 0    & Fraction of instantaneously recycled wind & (\ref{e:Mwindin}) \\
$\gaminf$  & 0.02 & Fraction of gas inflowing inside the disc per dynamical time & (\ref{e:Mmig_def}) \\
$\gamdis$  & 2    & Ratio of the turbulence dissipation timescale to the dynamical time & (\ref{e:gamdis}) \\
$\xi_i$    & 1    & Fraction of the in-streaming kinetic energy that turns into turbulence & (\ref{e:energy_steady_state}) \\
$\xi_m$    & 0    & Fraction of the disc inflow potential energy that turns into turbulence & (\ref{e:energy_steady_state}) \\
$u$        & $1/\sqrt{2}$ & Ratio of $\Vrot$ to the in-streaming velocity & (\ref{e:u}) \\
$\nu-1$    & $0$  & The slope of the logarithmic rotation curve, $(\log \Vrot/\log r)$ & (\ref{e:nu}) \\
\hline
\hline
\end{tabular}
\vspace{0.5cm}
\end{table*}

\subsection{Turbulent Energy steady state}
\label{s:energy_steady_state}
Gravitational energy can be transferred into turbulent energy to compensate
for the dissipative losses in two general ways. 
If a large fraction of the incoming streams that hit the disc is in dense gas 
clumps, they can transfer momentum into the disc gas, and thus convert the
stream kinetic energy into turbulence
\citep{GenzelR_08a,DekelA_09b,ElmegreenB_10a}.
Alternatively, the rotational and potential energy of the disc mass becomes
available due to the instability-driven mass inflow within the disc, including
clump migration,
where turbulence is generated by the same torques that are responsible
for angular-momentum outflow and the associated mass inflow
\citep{DekelA_09b,KrumholzM_10b,CacciatoM_11a,ForbesJ_12a}.
The quasi-steady state of the turbulent energy in the disc is then described by
\bea
\label{e:energy_steady_state}
\dot{K}_g&=&\xi_i (\dot{M}_{\rm cosmo}+\dot{M}_{\rm w,in})0.5V_{\rm in}^2+\xi_m\dot{M}_{\rm inf}\Vrot^2\\\nonumber
&-&(\dot{M}_{\rm SF}+\dot{M}_{\rm w,out}+\dot{M}_{\rm inf})1.5\sigma^2-M_{\rm g}1.5\sigma^2\tdis^{-1}=0,
\eea
where the different terms are as follows:
\begin{itemize}
\item
It is assumed that both the cosmological accretion and the recycled wind arrive to the disc at a speed
\be
V_{\rm in}\equiv \Vrot/u \, ,
\label{e:u}
\ee 
and that a fraction $\xi_i$ of the kinetic energy they carry is converted into disc turbulent energy. We take $u^2=0.5$ (see Appendix \ref{s:param_values}), and a fiducial maximum value $\xi_i=1$, so that we can examine the maximum possible contribution of in-streaming energy.
\item
It is assumed that as gas flows in or migrates to the galaxy centre, potential energy, which is released at a rate of $\sim\dot{M}_{\rm inf}\Vrot^2$, is transformed into turbulent energy with an efficiency $\xi_m$. See Appendix \ref{s:inflows_efficiency_appendix} for a discussion of this assumption. We later examine both the effects of $\xi_m=0$ and $\xi_m=1$.
\item
The gas sink terms, i.e.~star formation, outflowing winds and inflows to the bulge, take their share of turbulent energy when they leave the disc.
\item
The turbulent energy dissipates on a dissipation timescale $\tdis\equiv\gamdis\tdyn$, as in \equ{gamdis}, with a fiducial value $\gamdis=2$.
\end{itemize}

We solve \equ{energy_steady_state} for $\sigma/\Vrot$, using \equs{Mg} and (\ref{e:SFR}), and obtain 
\be
\frac{\sigma^2}{\Vrot^2}=\frac{2/3}{u^2/0.5}\frac{\xi_i(1+\etaw+\gaminf\epsSFR^{-1})+\xi_m \gaminf\epsSFR^{-1}(u^2/0.5)}{(\epsSFR\gamdis D)^{-1}+(1+\etaw+\gaminf\epsSFR^{-1})} \, .
\label{e:sigma2_full}
\ee
In the limit\footnote{With the fiducial values (Table \ref{t:parameters}), $1+\etaw+\gaminf\epsSFR^{-1}=3$ and $(\epsSFR\gamdis D)^{-1}=25$, so we consider the approximation $1+\etaw+\gaminf\epsSFR^{-1} \ll (\epsSFR\gamdis D)^{-1}$ a good one. More generally, $\etaw$ and $\gaminf\epsSFR^{-1}$ are not expected to have a value of more than a few, while $(\epsSFR\gamdis D)^{-1}$ is expected to have a value of several tens.} $1+\etaw+\gaminf\epsSFR^{-1} \ll (\epsSFR\gamdis D)^{-1}$ and with $u^2=0.5$, we obtain
\bea
\label{e:sigma2_limit}
\frac{\sigma}{\Vrot} &\approx& 0.16 \sqrt{\epsSFRII\gamdisII D}\\\nonumber
&\times&\sqrt{\xi_i(1+\etaw+\gaminf\epsSFR^{-1})+\xi_m \gaminf\epsSFR^{-1}},
\eea
where $\epsSFRII\equiv\epsSFR/0.02$ and $\gamdisII\equiv\gamdis/2$.

As already noted for the approximate steady-state solution in \equ{backbone_V_over_sigma_with_value}, the more detailed result for $\sigma/\Vrot$ in \equ{sigma2_full} is also independent of $\dot{M}_{\rm cosmo}$, because in our current model  the latter controls both the incoming energy and the disc gas mass that is involved in the turbulence, so the varying cosmological supply always provides the same {\it specific} turbulent energy.

\section{Case I: Non-regulated Disc Instability}
\label{s:caseI}
\subsection{The relative roles of in-streaming and disc inflow}
\label{s:cosmological_energy_discussion}
The approximate expression $\dot{E}_{\rm cosmo}\approx\dot{M}_{\rm cosmo}\Vrot^2$ used for the simple derivation in Section \ref{s:backbone} does not distinguish between cosmological in-streaming kinetic energy and gravitational potential energy released during inflows inside the disc. Each of these components carries similar specific energy of $\approx\Vrot^2$ (Appendix \ref{s:param_values}). Inspection of \equ{sigma2_limit} elucidates their individual roles.

The conversion of rotational and potential energy into turbulent energy depends on inflow in the disc, i.e.~$\xi_m$ couples to $\gaminf$. The ability of the kinetic energy of in-streaming mass to be converted into turbulence in the disc depends on complex physical processes in the vicinity of the disc, which we simply parameterize with $\xi_i$. If $\xi_i>0$, the accretion energy acts as a direct external driver of turbulence independently of disc inflows or winds. However, the presence of disc inflows and/or winds enhances the contribution of in-streaming in driving turbulence, i.e.~$\xi_i$ couples both to $\gaminf$ and $\etaw$. Thus, the relative contribution of kinetic and potential energy of cosmological origin depends not only on the intrinsic efficiencies $\xi_i$ and $\xi_m$, but also on the importance of winds and disc inflows.

\subsection{The effects of galactic winds and disc inflow}
\label{s:wind_effect}
By comparing \equs{Mg} and (\ref{e:backbone_Mg}) we can verify the role played by winds and inflows in the disc on the gas mass and hence the SFR. The gas mass and SFR are suppressed by escaping winds. In the limit of very strong winds and little recycling, $\etaw(1-\gamrcl)\gg 1+\gaminf\epsSFR^{-1}$, the gas mass and SFR are roughly proportional to $\etaw$. If the winds are fully recycled, $\gamrcl=1$, then they have no net effect on the gas mass and SFR, given that recycling is assumed to be instantaneous.

On the other hand, the value of $\sigma$ is independent of the recycling rate $\gamrcl$. This is despite the fact that the origin of the dependence on $\etaw$ does depend on $\gamrcl$. For example, If there is no recycling, $\gamrcl\approx0$, $\sigma$ is higher for larger $\etaw$ because the gas mass is smaller while the energy input is the same. On the other hand, if the outflows are all recycled back to the galaxy, $\gamrcl\approx1$, $\sigma$ is higher for larger $\etaw$ because the returning winds add to the energy provided by the cosmological accretion while the gas mass remains unchanged. We note that, regardless of the level of recycling, stronger outflows drive an increase in $\sigma$, despite the fact that we have ignored direct local deposit of feedback energy in the disc gas. This is done indirectly, either by lowering the gas mass or by adding to the energy brought by external accretion.

The inflow in the disc depletes the gas mass and thus suppresses the SFR, via $\gaminf\epsSFR^{-1}$, a quantity of order unity. The turbulent velocity $\sigma$ is even more sensitive to the inflow in the disc because (a) when the gas mass is depleted the same amount of energy input by in-streaming results in higher $\sigma$ (via $\xi_i$), and (b) the inflow down the potential gradient in the disc contributes energy to driving $\sigma$ up (via $\xi_m$).

Can the observed velocity dispersion at $z \sim 2$ be primarily driven
by the cosmological in-streaming?
Recall that the observed values are typically $\sigma/\Vrot \sim 0.2$, 
perhaps even $ \sim 0.3$.
If the conversion efficiency of in-streaming energy to turbulence is high, $\xi_i\approx1$,
and the SFR and dissipation rate are at their fiducial values,
the term referring to the streams by themselves already provides 
$\sigma/\Vrot \approx 0.16$, which
is in the ballpark of the desired value, though slightly short.
With the fiducial depletion by winds, $\etaw \sim 1$, 
this becomes $\sigma/\Vrot\gsim 0.2$. 
With the fiducial disc inflow, $\gaminf\epsSFR^{-1} \sim 1$ (and $\xi_m\approx1$), even without
winds, it becomes $\sigma/\Vrot \lsim 0.3$. 
Adding the fiducial winds and disc inflow, we obtain $\sigma/\Vrot \gsim 0.3$.
We conclude that if somehow $\xi_i \sim 1$, the gravitational energy associated
with the inflow, and in particular the clumpy gas streaming into the disc, 
can have a significant contribution to the disc turbulence, which can be
naturally aided by outflow depletion and disc inflow.

\subsection{Evolution of $Q$}
\label{s:Df1}
If the values of the parameters $D$, $\xi_i$, and $\xi_m$ are fixed, both $M_{\rm g}$ and $\sigma$ are determined by a balance between the externally set in-streaming, the star formation, the winds, and the inflows in the disc. In this case, the Toomre $Q$ parameter is not necessarily locked to $Q \sim 1$. Self-regulation of the instability at $Q \sim 1$ requires that the relevant physical processes, such as the inflow rate in the disc, the SFR and the outflow rate, adjust themselves to maintain $Q \sim 1$, and this case, where the relevant parameters are not fixed, is deferred to Section \ref{s:solutions_Q_1}.

The Toomre stability parameter is 
\be
Q = \frac{\sigma \kappa}{\pi\,G\Sigma_{\rm g}} \, ,
\label{e:Q1}
\ee
where $\kappa = \sqrt{2}\nu \Omega$. Using $\Omega=\Vrot/R$, $\Vrot^2=G M_{\rm tot}/R$, $M_{\rm tot}=M_{\rm g}/\delta$, and $M_{\rm g} = \pi\,R^2 \Sigma_{\rm g}$, we obtain
\be
Q = (2\nu)^{1/2} \delta^{-1} \frac{\sigma}{\Vrot} \, .
\label{e:Q2}
\ee
Then using the approximate solution in \equ{sigma2_limit} for $\sigma/\Vrot$, and assuming a flat rotation curve $\nu=1$, we obtain
\bea
\label{e:Q_2}
Q &\approx& 0.68 \delta_{0.33}^{-1} \sqrt{\epsSFRII\gamdisII D}\\\nonumber
&\times&\sqrt{\xi_i(1+\etaw+\gaminf\epsSFR^{-1})+\xi_m \gaminf\epsSFR^{-1}},
\eea
where the disc mass fraction (\equ{delta_def}) is expressed by $\delta_{0.33} \equiv \delta/0.33$. Substituting the fiducial values from Table \ref{t:parameters}, we obtain $Q\approx1.18\delta_{0.33}^{-1}$.

From \equ{Q2}, in the non-self-regulated case studied here, where $\sigma/\Vrot$ is constant, we learn that $Q$ scales with mass and time as $Q \prop \delta^{-1}$, where $\delta=M_{\g}/M_{\rm tot}$ within the disc radius. We can evaluate the time evolution of $Q$ through $\delta$ as follows. From \equ{Mg}, the steady-state solution is $M_{\rm g} \propto \dot{M}_{\rm cosmo} \tdyn$. Assuming that the disc radius is proportional to the halo virial radius, $R = \lambda R_{\rm vir}$, with $\lambda$ a constant spin parameter\footnote{This is based on the assumption of gas angular momentum conservation during the collapse to the disk \citep{MoH_98a}. While numerical studies have suggested that this picture neglects many details (e.g.~\citealp{ZavalaJ_08a,DuttonA_09a,DanovichM_12a,AumerM_12a}), it provides a fair match to observed disk sizes (e.g.~\citealp{MoH_98a,BullockJ_01a,MallerA_02a,DuttonA_07a,BurkertA_10a,DuttonA_12a}).}, we obtain $\tdyn \prop t_{\rm Hubble}$, which in the Einstein-deSitter phase that is approximately valid at $z>1$ gives
\be
\tdyn \prop (1+z)^{-3/2}.
\label{e:tdyn_vs_z}
\ee
Based on the EPS approximation (confirmed by fits to cosmological simulations), the cosmological input rate can be approximated in the Einstein-deSitter phase by
\be
\dot{M}_{\rm cosmo} \propto M_{\rm vir}\, (1+z)^{5/2}
\label{e:Mdot_cosmo}
\ee
\citep{NeisteinE_06a,NeisteinE_08b}. Assuming that the halo mass profile is roughly $M(r) \prop r$, we have $M_{\rm tot} \simeq \lambda M_{\rm vir}$\footnote{This is formally exact for an isothermal sphere, which is a reasonable approximation for the dark matter halo. Since typically $\lambda\sim0.03$ \citep{BullockJ_01a}, and the baryons in the central galaxy are known to be less massive than $\approx0.03M_{\rm vir}$ \citep{GuoQ_10a,BehrooziP_10a}, we find that this is a reasonable approximation altogether.}. The above finally yield for the non-self-regulated model
\be
Q \propto \delta^{-1} \propto (1+z)^{-1} \, .
\label{e:Q_scaling}
\ee
We learn from \equs{Q_2} and (\ref{e:Q_scaling}) that the discs tend to be unstable at high redshifts, $Q < 1$, and then evolve toward stabilisation at later times, $Q > 1$. We can see that the growth of $Q$ in time is driven by the decline of the cosmological accretion rate, \equ{Mdot_cosmo}, being steeper than the increase in time of $\tdyn$. This trend is in at least qualitative agreement with observations that find $Q$ to be of order a few in local disc galaxies (e.g.~\citealp{LeroyA_08a,WatsonL_12a}), but around, or even below, unity at high redshift \citep{GenzelR_11a}.

\Equ{Q_2} shows that at high redshift, violent disc instability with $Q\lesssim1$ is naturally 
driven by the high $\delta$, which results from the intense in-streaming rate, \equ{Mg}.
The in-streaming power, even when the efficiency for driving
turbulence is high, $\xi_i=1$, may not be able to drive turbulence with
$\sigma/\Vrot$ high enough for balancing the high $\delta$ and 
thus stabilising the disc. 
The instability is associated with disc inflow of a large 
$\gaminf$, which can drive further turbulence both by providing energy that 
is converted to turbulence if $\xi_m$ is high and by depleting 
the disc mass (see Section \ref{s:m}). 
High-redshift observations reveal unstable discs where $Q \lsim
1$, with a high $\delta$ \citep{TacconiL_10a} as well as a high $\etaw$
(e.g.~\citealp{SteidelC_10a,GenzelR_11a}). The latter may indicate that 
feedback also plays a role in maintaining $Q \sim 1$, either indirectly by 
depletion of the disc gas through outflows,
or by direct injection of feedback energy into turbulence (not
considered here). 
Some combination of disc inflows and feedback/outflows is capable of providing turbulence
at the level that would stabilize the disc, and this may occur in a self-regulated way
that keeps the disc marginally unstable.
At low redshift, where $\delta\ll1$, $Q$ could be higher than
unity even without the presence of winds and disc inflows (see also
our complementary calculation in \citealp{CacciatoM_11a}).

We can replace the assumption of a constant loading factor $\etaw$ with a prediction for momentum-driven winds, $\etaw \propto V^{-1}$ \citep{MurrayN_05a}. Using the standard cosmological virial relation this implies 
\be
\etaw \propto M^{-1/3} (1+z)^{-1/2} \, .  
\label{e:eta}
\ee
Plugging this in \equ{sigma2_limit}, in the limit of strong outflows $\etaw \gg 1$, we obtain
\be
\frac{\sigma}{\Vrot} \propto \etaw^{1/2} \propto \Vrot^{-1/2} \propto M^{-1/6} (1+z)^{-1/4} \, .
\label{e:sigma_over_V_eta_scaling}
\ee
Then from \equ{Q2}, using \equ{Mg} that implies $\delta \propto \eta^{-1} (1+z)$, we obtain\footnote{Note 
that there could be tension between the limit
$1+\etaw+\gaminf\epsSFR^{-1} \ll (\epsSFR\gamdis D)^{-1}$ in which
\equ{sigma2_limit} is valid, and the limit $\etaw \gg 1$ in which
\equ{sigma_over_V_eta_scaling} is valid.
They may both be valid if $\etaw \sim 5$, say.
If $\etaw$ is larger, we can relax the first condition 
and use \equ{sigma2_full} instead of \equ{sigma2_limit}.
We then find that the scaling of $Q$ in the $\etaw \gg 1$ limit is bounded by
the scalings in \equs{Q_scaling} and (\ref{e:Q_eta_scaling}),
and the scaling of $\sigma/\Vrot$ is similarly bounded by the flat
value of \equ{sigma2_full} and the weak variation in
\equ{sigma_over_V_eta_scaling}.} 
\be
Q \propto M^{-1/2} (1+z)^{-7/4} \, .
\label{e:Q_eta_scaling}
\ee
We learn that the variation of $\etaw$ according to momentum-driven winds enhances the redshift dependence of $Q$ compared to the constant $\etaw$ case, \equ{Q_scaling}. Unfortunately, the slight growth of $\sigma/\Vrot$ in time in \equ{sigma_over_V_eta_scaling} makes the agreement with the observed {\it decline} of $\sigma/\Vrot$ even worse than in the constant $\etaw$ case. The consideration of momentum-driven winds also introduces a mass dependence. The predicted weak mass dependence of $\sigma/\Vrot\propto(1+z)^{-1/4}$ seems reasonable, although the observational trend is not yet well established. For example, \citet{KlessenR_10a} find that $\sigma/\Vrot \propto M^{-1/3}$ in the local Universe, while \citet{vanderKruitP_11a} report $\sigma/\Vrot \approx$const. At $z\sim2$, the largest existing observational sample shows no clear trend, though still with significant scatter that makes the situation inconclusive \citep{ManciniC_11a}.

The observed mass dependence of $Q$ in galaxy discs is not yet well established. While \citet{DalcantonJ_04a} find a sharp threshold for the onset of instability at $\Vrot>120\kms$, \citet{WatsonL_12a} find no trend of $Q$ with galaxy mass. Our predicted mass dependence of $Q$ depends on the scaling of $\delta$ and the different model parameters with mass. In the case of constant model parameters, we obtain that $Q$ is independent of mass, \equ{Q_scaling}, and in the case of momentum-driven winds scaling, we obtain that more massive discs are less stable, \equ{Q_eta_scaling}.

\section{Case II: Self-Regulated Disc Instability}
\label{s:solutions_Q_1}
In Section \ref{s:caseI} we have shown that case I produces gas discs in which $\sigma/\Vrot$ is constant in time and $Q$ is gradually increasing to values larger than unity. While the systematic increase in $Q$ toward low redshifts may be consistent with the observational trend toward a larger abundance of stable discs, the constancy of $\sigma/\Vrot$ is clearly in conflict with the observed decline of this quantity. In this section, we appeal to an alternative case II, 
where self-regulation at $Q \sim 1$ is imposed at all times.
We address the possibility that the self-regulation is achieved by periodic 
episodes where the instability and the associated SFR and disc inflows are `on' or `off' 
with a duty cycle $D<1$ (see also \citealp{MartigM_09a}). Self-regulation may alternatively be achieved 
by adjustments of the disc inflow rate, via $\gaminf$ and $\xi_m$,
as in our companion paper \citep{CacciatoM_11a}. 
We also check the effect of a systematic decline with time of the accretion 
conversion efficiency $\xi_i$, perhaps reflecting the evolution from dense narrow
streams to a wide-angle accretion and a gradual decrease in the gas fraction 
and degree of clumpiness in the accreting gas.

\subsection{Maximum conversion efficiencies and a duty cycle}
\label{s:D}
We assume here that once $Q$ rises to slightly above unity, the disc
tends toward stabilisation, star formation is suppressed \citep{KennicuttR_89a,MartinC_01a}, 
and the galactic outflow as well as inflow within the disc become weaker too. This allows $\Sigma_{\rm g}$ to increase in response to the continuing cosmological accretion, while the growth of $\sigma$ slows down.
As a result, $Q$ tends to decrease to slightly below unity, back to an 
unstable phase where star formation, outflows and disc inflow resume, and so on.
We thus expect oscillations about $Q\sim1$. 
We model this by a duty cycle $D<1$ that represents the fraction of time when 
the instability is `on'. 
Here we adopt maximum efficiencies for energy conversion to turbulence, 
$\xi_i=\xi_m=1$.
From \equ{Q_2}, with $Q=1$, we obtain
\be
D \simeq \frac{2.2 \delta_{0.33}^2}
{\epsSFRII\gamdisII(1+\etaw+2\gaminf\epsSFR^{-1})} \, .
\label{e:D_2}
\ee

With the fiducial parameter values (Table \ref{t:parameters}), and a high cold mass fraction
characteristic of high-redshift discs, $\delta \simeq 0.33$,
we have $D \lsim 1$.
When the gas fraction is even higher, the outflow rate is lower, 
or the disc inflow
rate is lower, \equ{D_2} may give $D>1$, which is clearly unphysical.
This is another representation of the result from Section \ref{s:Df1}
that at high redshift, when $\delta$ is high, in the absence of winds and
when the disc inflow is ignored, there is hardly enough power in the
cosmological in-streaming by itself to drive the high turbulence required for $Q=1$.
The fiducial winds, $\etaw\sim 1$, or the fiducial disc inflow,
$\gaminf\epsSFR^{-1}\sim 1$, help obtaining a physical result with $D<1$.
Another way to obtain $D<1$ may be if
during the violent instability phase the timescales for star formation and
inflow somehow become shorter than the disc dynamical time $\tdyn$, while the
dissipation timescale is long compared to $\tdyn$.

On the other hand, at low redshift, with $\delta\lesssim0.1$,
and even more so if $\gaminf$ and $\etaw$ are non-negligible, 
the resulting duty cycle $D$ drops significantly below unity to $\lsim0.1$
in order to keep $Q=1$. From \equ{D_2}, the scaling of $D$ with mass and redshift
is $D\propto M^0(1+z)^{2/3}$ in the limiting case of weak winds ($\etaw \ll 1$),
and $D\propto M^{1/3}(1+z)^{7/6}$ in the limit of 
strong momentum-driven winds with no recycling ($\etaw\gg1$ and $\gamrcl=0$). In the general
case between these limits, $D$ is declining with time close to linearly with
$(1+z)$, such that at later times one expects to detect a smaller fraction of the
galaxies in the unstable phase.

When forcing $Q=1$ and keeping $D$ free, from \equs{Mg} and (\ref{e:sigma2_limit}) and the definitions for $\kappa$ and $\Sigma_{\rm g}$, we obtain
\bea
\label{e:Sigma_D}
\Sigma_{\rm g} &=& \frac{\kappa}{\pi}\left(\frac{\gamdis\dot{M}_{\rm cosmo}(1+\etaw+2\gaminf\epsSFR^{-1})}{3\sqrt{2}u^2G^2(1+\etaw(1-\gamrcl)+\gaminf\epsSFR^{-1})}\right)^\frac{1}{3}\\\nonumber
&\approx& 36\Msunpc2\left(\frac{35\Myr}{\kappa^{-1}}\right)\left(\frac{\dot{M}_{\rm cosmo}}{1\Msunyr}\right)^\frac{1}{3}
\eea
and 
\bea
\label{e:sigma_D}
\sigma &=& \left(\frac{G\dot{M}_{\rm cosmo} \gamdis(1+\etaw+2\gaminf\epsSFR^{-1})}{3\sqrt{2}u^2(1+\etaw(1-\gamrcl)+\gaminf\epsSFR^{-1})}\right)^\frac{1}{3}\\\nonumber
&\approx& 17.5\kms\left(\frac{\dot{M}_{\rm cosmo}}{1\Msunyr}\right)^\frac{1}{3},
\eea
where the second equality in each of these equations is calculated using our fiducial parameters from Table \ref{t:parameters}.
In contrast with the solution in Section \ref{s:caseI}, here
$\sigma/\Vrot$ does depend on the external cosmological in-streaming rate 
$\dot{M}_{\rm cosmo}$. This is a result of the imposed self-regulation, where a decrease
in $\dot{M}_{\rm cosmo}$ induces similar decreases 
in $\sigma/\Vrot$ and in $\delta$ such that $Q$ remains constant, \equ{Q2}.
These evolution trends agree with the observed trends better than those predicted in case I. In particular, the results obtained for $\dot{M}_{\rm cosmo}=1\Msunyr$ and $\kappa=(35\Myr)^{-1}$, as roughly appropriate for the Milky Way, are close to the typical values in local disc galaxies, and if $\dot{M}_{\rm cosmo}$ is scaled up to $\approx100\Msunyr$, as appropriate for high redshift discs, one obtains $\Sigma_{\rm g}\approx200\Msunpc2$ and $\sigma\approx80\kms$, which are indeed in the observed ballpark \citep{FoersterSchreiberN_09a,GenzelR_10a}.
In case II the gas mass $M_{\rm g}$ declines at a slower rate,
because the overall gas depletion by star formation, outflows and disc inflow
is slower in accord with the low duty cycle $D$.
An examination of \equ{Mg} using the aforementioned scaling of $D$ in this solution, as well as \equs{tdyn_vs_z} and (\ref{e:Mdot_cosmo}), gives $M_{\rm g}\propto M(1+z)^{{1}/{3}}$. Similarly, \equs{Sigma_D} and (\ref{e:sigma_D}) give $\Sigma_{\rm g}\propto M^{{1}/{3}}(1+z)^{{7}/{3}}$ and $\sigma\propto M^{{1}/{3}}(1+z)^{{5}/{6}}$, respectively, and $\sigma/\Vrot\propto M^0(1+z)^{{1}/{3}}$, for any constant $\etaw$.
These trends with redshift agree reasonably well with observations as well. 
The associated evolution rate of the gas fraction, 
$M_{\rm g}/M\propto (1+z)^{{1}/{3}}$, is however somewhat weaker than
observations indicate for massive galaxies between $z\approx2$ and $z=0$
(see \citet{BoucheN_10a} and references therein).

We note, however, that a decreasing star formation duty cycle, in particular one that decreases as strongly as $\delta^2$, should manifest itself in observations. If the `off' periods are long compared to the galactic dynamical time, $D\ll1$ will result in a large fraction ($\approx1-D$) of disc galaxies having significant gas discs with no star-formation. There are indeed several examples of observed gas discs that have low levels of star-formation, possibly as a result of their stability (e.g.~\citealp{MartigM_09a,MacLachlanJ_11a}). However, for typical disc galaxies this is not the case. Indeed, \citet{LeroyA_08a} and \citet{WatsonL_12a} find no clear correlation between star-formation efficiency and $Q$. On the other hand, if the `off' periods are short compared to the galactic dynamical time, they will result in discs being in a mixed state of `on' and `off' in different regions on the disc. In a statistical sense, this will show in observations as a decrease of the mean star-formation efficiency with 
cosmic time. Such a trend may be observed, but probably not to the extent that is predicted by our model \citep{GenzelR_10a,DaddiE_10b,KrumholzM_12a}. Disc galaxies are observed to have, on average, similar star-formation efficiencies per dynamical time at low and high redshift. This remains an open problem with regards to this scenario.

\subsection{Sub-maximal efficiencies and no duty cycle}
\label{s:low_conversion}
We now allow only a fraction $\xi_i$ of the incoming kinetic energy to be
converted into turbulent energy, and solve for the value of $\xi_i$ that is 
required to keep $Q=1$. We start with the accretion energy as the only
direct driver of turbulence, namely $\xi_m=0$, and assume $D=1$. 
From \equ{Q_2}, with $Q=1$, we obtain
\be
\xi_i \simeq \frac{2.2 \delta_{0.33}^2}
{\epsSFRII\gamdisII(1+\etaw+\gaminf\epsSFR^{-1})} \, .
\label{e:xim0_xii_2}
\ee
This is a very similar expression to the one found for $D$ in \equ{D_2}. However, the physical situation is different in the two cases. We note that $\delta$ evolves differently because in the $D\neq1$ case $M_{\rm g}$ is larger than in the current case, as shown in Section \ref{s:D}. In the current solution, $\Sigma_{\rm g}$ is the same as in case I, i.e.~it is decreasing in proportion to $\dot{M}_{\rm cosmo}$, but due to the requirement $Q=1$, here $\sigma$ is forced to decrease as well. This is achieved by a decreasing conversion efficiency $\xi_i$ from in-streaming to turbulence.

The alternative, more physically motivated assumption of efficient conversion
of the energy associated with the disc inflow, $\xi_m=1$
(Appendix \ref{s:inflows_efficiency_appendix}), leads to
\be
\xi_i \simeq \frac{2.2 \delta_{0.33}^2\gamdisII^{-1}-50\gaminf}
{\epsSFRII(1+\etaw+\gaminf\epsSFR^{-1})} \, .
\label{e:xim1_xii_2}
\ee

In both of these $\xi_i<1$ cases, the scalings of $M_{\rm g}$, $\delta$ and $\Sigma_{\rm g}$ with mass and time are the same as in case I. From $Q=1$ we obtain $\sigma/\Vrot=\delta/\sqrt{2\nu}$, which implies
\be
\sigma/\Vrot\propto M^0(1+z)^1
\label{e:sigma_over_V_scaling_with_efficiency}
\ee
without winds, and
\be
\sigma/\Vrot\propto M^{-\frac{1}{3}}(1+z)^{1.5}
\label{e:sigma_over_V_scaling_with_efficiency_winds}
\ee
with escaping strong winds. The evolution of $\sigma/\Vrot$ in the cases where $\xi_i<1$, with or without winds, agrees better with observations.

At lower redshift, as the cold disc fraction $\delta$ declines, 
the condition $Q=1$ with $D=1$ forces the conversion efficiency of the 
in-streaming to turbulence $\xi_i$ to decline as well, $\xi_i\propto\delta^2$,
as in \equs{xim0_xii_2} and (\ref{e:xim1_xii_2}). This means $\xi_i\propto (1+z)^2$
without winds and $\xi_i\propto M(1+z)^{3.5}$ with escaping strong winds.
There could be several reasons for such a decline of $\xi_i$ in time.
First, the conversion factor could be a growing function of the gas fraction in 
the disc, namely of $\delta$.
Second, it is likely to be growing with the gas fraction in the accreting
baryons, which is declining with time.
Third, the evolution of the input pattern from narrow, dense streams at high
redshift to a wide-angle accretion at late times makes the penetration
of cold gas into the disc less efficient at later times \citep{DekelA_06a,vandeVoort_11a}.
The lower gas density contrast between 
the streams and the disc is another reason for a smaller $\xi_i$, 
especially if it is somehow associated with a lower degree of clumpiness in the
streams, as the coupling between in-streaming and disc arises from in-streaming gas of 
comparable density to the disc gas \citep{DekelA_09b}.
However, except for the first reason that relates to the gas fraction in the
disc, the value of $\xi_i$ is determined externally,
independently of the instability state of the disc or its other properties. 
Therefore a self-regulation loop is not expected to drive the required 
$\xi_i \propto \delta^2$ for $Q \sim 1$.
In this case, it seems that the evolution of $\xi_i$ can only match the 
requirements for $Q \sim 1$ by some coincidence.

\subsection{Self-regulated inflow in the disc}
\label{s:m}
Assuming fixed values for $\xi_i$, $\xi_m$ and $D$, the inflow rate
within the disc may adjust itself to compensate for the dissipative losses
and maintain the instability at $Q \sim 1$.
This is the basis of the analysis in
\citet{KrumholzM_10b}, \citet{ForbesJ_12a}, and our companion paper
\citet{CacciatoM_11a}. 
With our fiducial choice of parameters, including outflows, this can be
achieved quite naturally.
In the absence of outflows, the disc inflow rate should be comparable to
or somewhat higher than the SFR, $\gaminf\gtrsim\epsSFR$, 
depending on the exact values of the parameters $\xi_i$, $\xi_m$ ,$\gamdis$ and 
$\epsSFR$ (see a discussion of \citet{KrumholzM_10b} in Section 
\ref{s:literature}). 
We note that an enhanced instability that boosts up the disc inflow rate
would also enhance the SFR and outflow rate, which would help the
self-regulation. However, if such a case is also accompanied by faster dissipation,
then increased inflows may not be able to solve the problem.
The details of the inflow within the disc, including gas and stars, clumps
and off-clump material, is being investigated via cosmological hydrodynamical
simulations (Cacciato et al., in preparation). 
This has additional important consequences regarding the issues of bulge 
growth and feeding central black holes \citep{BournaudF_11a,BournaudF_12a}.

\section{Discussion: comparison with the literature}
\label{s:literature}
Several earlier studies evaluated the possible role of in-streaming and disc inflows
in driving the observed velocity dispersion in $z\sim2$ discs, reaching seemingly
conflicting conclusions. The first estimates of this kind 
\citep{FoersterSchreiberN_06a,GenzelR_08a} suggested that it is plausible that
there is enough energy in the cosmological in-streaming, depending on the exact
numerical values of several parameters in their equations. In
\citet{DekelA_09b} the in-streaming energy explicitly depends on the 
unknown small-scale clumpiness of the streams, and could therefore go either 
way. In \citet{CacciatoM_11a} we obtain 
$\sigma/\Vrot\approx0.2$ when choosing favourable values for the relevant 
parameters. \citet{KlessenR_10a} estimated that there is enough energy in the
cosmological accretion, and \citet{KrumholzM_10b} reached a similar conclusion 
when examining the gravitational potential energy that is released during mass
inflows inside the discs. \citet{KhochfarS_09a} concluded that only $18\%$ of the in-streaming energy is required to reproduce the observed values. In apparent contrast, \citet{LehnertM_09a} and
\citet{ElmegreenB_10a} conclude that there is not enough in-streaming energy to account for $z\sim2$ turbulence.

As we show in Section \ref{s:conservation_eqs}, the value of $\sigma/\Vrot$ depends basically on two timescales, one associated with the decay of turbulence (parameterized with $\gamdis$), and the other associated with the gas mass conservation in the disc (related to $\epsSFR$, $\etaw$ and $\gaminf$). All of the aforementioned studies involved a turbulent energy balance that is very similar to the one we consider in this work. However, they differ from one another in the choices of the turbulent dissipation timescale. Moreover, some of these models do not explicitly include a mass steady state condition, and differ in the {\it implicit} assumptions they make regarding the timescale associated with the mass equation. A closer inspection of the different assumptions made, which we perform next, reveals how they lead to the apparently conflicting conclusions.

\citet{GenzelR_08a} do not explicitly write a steady state equation for the gas mass conservation, but their treatment is equivalent to the simple equations in Section \ref{s:conservation_eqs}, only with $\tsf$ crudely approximated by the specific rate of cosmological accretion $t_{\rm acc}$ (from the growth of dark matter haloes), hence not addressing the possible difference between the rates of star formation and accretion. They remain agnostic as for whether there is enough in-streaming energy for driving the observed $\sigma/\Vrot\approx0.2-0.3$, due to uncertainties in numerical values of order unity. Nevertheless, their fiducial values\footnote{Note that the factor $\beta^{-1}\gamma^{-2}$ in eq.~(10) of \citet{GenzelR_08a} should be corrected to $\sqrt{\beta^{-1}\gamma^{-2}}$ (Genzel, R., private communication).} indicate a small value of $\sigma/\Vrot\approx0.07$, implying that an additional source of energy is required. Apart from the choice of parameter values, there is a significant difference between our models. As $t_{\rm acc}$ declines with time faster than the dynamical time, \citet{GenzelR_08a} naturally obtain that $\sigma/\Vrot$ decreases with time (see \equ{backbone_V_over_sigma}). In our model, $M_{\rm g}$ follows $\dot{M}_{\rm cosmo}$ (\equ{Mg}) and becomes smaller with time, so that the effective timescale for specific mass and energy gain is constant (and related to $\tdyn$). With the \citet{GenzelR_08a} implicit assumption $\tsf=t_{\rm acc}$, the gas mass remains constant with redshift, since the star-formation timescale becomes longer at the same rate that the accretion diminishes.

\citet{KhochfarS_09a} find that a conversion efficiency of only $\xi_i=18\%$ is required to obtain high $\sigma/\Vrot$ as observed. Their semi-analytical model does not explicitly include a steady-state solution to the gas disc mass as we do here, but gas velocity dispersions are obtained in their model as a result of similar physical considerations to the ones made here in Section \ref{s:backbone}. Therefore, it is straight-forward to compare the two analyses and find that they require a lower conversion efficiency $\xi_i$ as a result of the higher star-formation efficiency $\epsSFR$ ($\alpha$ in their notation) they assume for their porosity-driven star-formation model, $\epsSFR\approx0.02(\sigma/10\kms)\approx0.15$ at $z\sim2$.

The set of equations developed in \citet{ElmegreenB_10a} is very similar to our equations here. They do not consider the winds ($\etaw$) and disc inflow ($\gaminf$) terms, while they do consider a constant conversion efficiency $\xi_i$ ($\epsilon$ in their notation) from in-streaming kinetic energy to turbulent energy and no star formation when $Q>1$. \citet{ElmegreenB_10a} focus on the transient state before the quasi-steady state, an issue we do not address here. Nevertheless, they obtain that once the quasi-steady state is reached, the accretion cannot be the dominant source of turbulence, because it cannot provide enough turbulence to keep the disc at $Q\sim1$. When we neglect winds and disc inflows, we arrive at a similar conclusion, unless we choose favourable parameter values such as $\epsSFR=0.03$ and $\gamdis=3$, for which we obtain $\sigma/\Vrot=0.24$ (see \equ{sigma2_limit}). An additional difference between the \citet{ElmegreenB_10a} work and ours is that they take the dissipation timescale to be the perpendicular crossing time, such that $\gamdis\equiv t_{\rm perp}\omega=(H/\sigma)(\pi G \Sigma_{\rm g}/\sigma)=(\sigma^2/\sigma\pi G \Sigma_{\rm disc})(\pi G \Sigma_{\rm g}/\sigma)=\delta$, while we assume $\gamdis$ to be a constant $\approx1-3$. As a result, their solution at low accretion rates (and low $\delta$) is not as over-stabilized as ours (but still more stable than a case with a higher accretion rate). However, they do not directly discuss the case where the accretion is low enough that the disc is over-stabilized at late times, where we have shown that a decreasing $\xi_i$ or $D$ may allow for a late marginally-unstable configuration.

\citet{KlessenR_10a} conclude that there is more than enough accretion energy to drive the observed turbulence at high redshift. They make a simple comparison between the energy input rate $0.5\dot{M}_{in}V_{\rm in}^2$ and the dissipation rate, and apply relevant numerical values in different situations to see which one is larger. In the case of clumpy discs at high redshift, it seems that they used unrealistic numerical values. Using appropriate values in their eq.~(8), either for individual clumps or for the whole disc, yields that a high conversion efficiency is required, $\xi_i\approx1$, rather than $\xi_i\ll1$ according to their estimate. This would imply that there is barely enough energy in the in-streaming for being the sole driver of the turbulence, in agreement with our results. In particular, for the case of single clumps, the result in their eq.~(16) should be compared with the SFR in a single $10^8\Msun$ clump, i.e.~$\sim1\Msunyr$, not with that of the whole disc, $10-50\Msunyr$. Such a proper comparison would give a conversion efficiency of $\approx1$. For the case of the whole disc, their choice of $\sigma_{\rm 3D}=30\kms$ is too low in the sense that it corresponds to $V/\sigma\gtrsim10$. A choice of $V/\sigma\approx5$ would require an efficiency that is $\approx10$ times higher than their estimate. In addition, there is a factor of $\approx3.2$ missing in the transition from their eq.~(15) to eq.~(16), since $(30\kms)^3/(2G)\approx3.2\Msunyr\neq1\Msunyr$.

\citet{KrumholzM_10b} and \citet{ForbesJ_12a} solve the evolution equations for a thin axisymmetric disc under the assumption of self-regulated marginal instability. The turbulent energy input rate is driven in their model by the inflow rate within the disc (corresponding to a fixed $\xi_m$ in our model), and it is dissipated on a crossing timescale. They solve the evolution equations for the required inflow rate to keep $Q=1$, which is very similar to the possibility of letting $\gaminf$ change in our model \citep[and as assumed in][]{CacciatoM_11a}, keeping all other factors fixed (Section \ref{s:m}). The \citet{KrumholzM_10b} solution is very similar to ours in case II where we force $Q=1$ (Section \ref{s:D}). However, in their fiducial solution for $z\approx2$ discs they obtain (see their eq.~(45)) an inflow rate that has to be roughly six times higher than the star-formation rate, or in our formalism $\gaminf/\epsSFR\approx6$. Similarly, the resulting disc inflow rate at $z\approx2$ in the \citet{ForbesJ_12a} fiducial disc, where they also assume outflows with $\etaw=1$, is higher than the SFR in disc (Forbes, J., private communication). This is another manifestation of our conclusion that the high turbulence at high redshift cannot be solely driven by the direct effect of the incoming streams, and it requires additional contributions from an intense inflow within the disc, and possibly a depletion by outflows. We conclude that the \citet{KrumholzM_10b} and \citet{ForbesJ_12a} results also agree with the other studies reviewed in this section.

\section{Summary and conclusions}
\label{s:summary}
We have developed an analytic model that describes certain aspects of the
evolution of galactic gas discs in a cosmological context. 
Specifically, we addressed the possibility that the streaming of
external gas into the discs, which provides new fuel for disc instability and
star formation, also directly drives turbulent motions by converting the
in-streaming kinetic energy into disc turbulence.
The results of the model agree with the observations of gas-rich $z\sim2$ galaxies,
where $\sigma/\Vrot\sim0.2-0.3$ and $Q\approx1$.
In the absence of self-regulation, with all the parameters fixed,
we find that the ratio of turbulent to rotation velocity
$\sigma/\Vrot$ remains constant in time independently of the accretion
rate, and that discs evolve from violent instability at high
redshift toward stability at low redshift. The constancy of
$\sigma/\Vrot$ with redshift does not agree with the trend suggested by observations. 
In a second case, where the discs are assumed to be self-regulated to marginal 
instability about $Q\approx1$ as a result of a duty cycle in the
instability and the associated star formation and outflows,
$\sigma/\Vrot$ is found to decline with cosmic time in better accord with the 
observations. In this case the duty cycle declines at late times
toward long periods of stability separating short episodes of instability.
Since there is no evidence at low redshift for a strong increase in the mean 
timescale for gas consumption into stars, the small duty cycle seems to be observationally unfavourable.

At $z\sim 2$, we conclude that only if the conversion efficiency of in-streaming
kinetic energy to turbulence is high, i.e.~close to unity, the direct driving of turbulence by 
the external accretion can be sufficient. Still, unless the SFR efficiency is on the high side of the common estimates, or the turbulence decay rate is much slower than the dynamical time, this scenario relies on disc depletion caused by either galactic winds or inflows within the disc, or both.
At low redshift, the driving of turbulence by the inflowing gas may actually be too much if the conversion efficiency remains high and the instability and disc depletion are assumed to be continuously `on' (see similar conclusions in \citealp{DekelA_09b}). A conversion efficiency or a duty cycle that decline with time may help recover the evolutionary trends.

We wish to emphasize the role played by the depletion from the disc in allowing a high velocity dispersion in the remaining disc gas. This depletion is a natural result of star formation, galactic winds, and inflows within the disc. With less gas in the disc, the same energy input would drive a higher velocity dispersion. In particular, if stellar feedback and/or AGN feedback drive massive outflows, the disc turbulence will be enhanced even if there is no direct energy injection from the feedback source into the remaining ISM. If some fraction of the outflows is recycled back into the disc, this may add to the direct stirring up of turbulence by accretion.

In a companion paper \citep{CacciatoM_11a}, we neglect the direct contribution of the in-streaming energy, assuming that the disc turbulence is powered by the inflow within the disc, which is intimately coupled to the self-regulation of the disc instability at $Q=1$. We find there that the instability is unavoidable at high redshift, because of the intense accretion that maintains a high gas fraction, and that the discs are driven to $Q>1$ at low redshift, primarily due to the growing dominance of the stellar component. In this model $\sigma/\Vrot$ tends to decline at late times. This is in qualitative
agreement with the trends we find here when we impose $Q \sim 1$.

In \citet{CacciatoM_11a} we considered both gas and stars in a two-component disc instability analysis, while here we limited the analysis to a one-component gas disc, in order to allow for an analytic solution. Our results there showed that in order to maintain marginal instability, the presence of a `hot' stellar component implies a lower gas velocity dispersion than in the one-component case. This indicates that the inclusion of stars in the model considered here would have made it even harder to properly suppress the turbulence driven by external accretion and reproduce marginal instability at low redshifts.

To summarize, our model suggests that turbulence driven by cosmological in-streaming may account for the high turbulence observed in $z\sim2$ discs, but only if the coupling between this inflow and the disc is high. On the other hand, at low redshift our model is in tension with observations unless the conversion efficiency of the in-streaming energy evolves in a certain way to a low value. Thus, cosmological in-streaming could in principle have an important role in driving turbulence in galactic disks, but for this to be the primary driver, and to hold throughout cosmic time, the energy conversion efficiency between inflow and disk has to be fine-tuned.

\section*{Acknowledgments}
We acknowledge stimulating discussions with Nicolas Bouch{\'e}, Andreas Burkert, John Forbes, Mark Krumholz, Amiel Sternberg, and Romain Teyssier.
This work has been supported by the
ISF through grant 6/08, by GIF through grant
G-1052-104.7/2009, by a DIP grant,
and by an NSF grant AST-1010033 at UCSC.
MC has been supported at HU by a Minerva fellowship (Max-Planck Gesellschaft).

\appendix
\section{Conditions for a quasi-steady state solution}
\label{s:quasi_steady_state_validity}
For the solution of \equ{mass_conservation} with the assumption $\dot{M}_{\rm sink}=M_{\rm g}\tau^{-1}$ to be \equs{mass_conservation_solution_Mgas} and (\ref{e:mass_conservation_solution}), it is required that $\dot{M}_{\rm cosmo}$ can be treated as a constant, i.e.~that it changes slowly with respect to the typical timescale of the solution, namely
\be
\left|\frac{d\dot{M}_{\rm cosmo}/dt}{\dot{M}_{\rm cosmo}}\tau\right|<1.
\label{e:timescales_comparison}
\ee
We can write
\be
\frac{d\dot{M}_{\rm cosmo}/dt}{\dot{M}_{\rm cosmo}}=\dot{M}_{\rm cosmo}^{-1}(\frac{\partial\dot{M}_{\rm cosmo}}{\partial t}+\frac{\partial\dot{M}_{\rm cosmo}}{\partial M}\frac{dM}{dt})
\label{e:Mcosmo_evolution}
\ee
since $\dot{M}_{\rm cosmo}$ changes with cosmic time for a given halo mass, while it also changes for a given halo as it builds up its mass.

For \LCDM{ }cosmology we can write approximately $\dot{M}_{\rm cosmo}\propto(1+z)^{2.2}\propto t^{-1.7}$, where $t$ is cosmic time \citep{NeisteinE_06a,GenelS_08a}. Therefore, the first term equals $\dot{M}_{\rm cosmo}^{-1}\partial\dot{M}_{\rm cosmo}/\partial t\approx-1.7t^{-1}$. Further, we can replace $dM/dt$ in the second term with $\dot{M}_{\rm cosmo}$, and define $t_{\rm acc}\equiv\partial\dot{M}_{\rm cosmo}/\partial M$ as in \citet{BoucheN_10a}, to obtain from \equ{timescales_comparison} the condition
\be
\tau^{-1}>\left|-1.7t^{-1}+t_{\rm acc}^{-1}\right|.
\label{e:final_Mcosmo_condition}
\ee

Since the two terms on the right-hand side have opposite signs, considering each of them separately gives the most stringent conditions that are required for $\dot{M}_{\rm cosmo}$ to be considered a constant when solving \equ{mass_conservation}. When we consider only the growth of $\dot{M}_{\rm cosmo}$ that accompanies the halo growth, \equ{final_Mcosmo_condition} becomes $t_{\rm acc}>\tau$. This condition is discussed in \citet{BoucheN_10a}, where it is found to hold for $z\lesssim7$. When, instead, we consider only the direct dependence of $\dot{M}_{\rm cosmo}$ on cosmic time, a stronger condition is obtained, as \equ{final_Mcosmo_condition} becomes $t>1.7\tau$. This is similar to the condition $t\gg\tau$ that is required to obtain the solution in \equs{mass_conservation_asymptotic_solution_Msink}-(\ref{e:mass_conservation_asymptotic_solution_Mg}).

Therefore, we only have left to consider the relation between $t$ and $\tau$. Since $\tau$ is shorter than the star-formation timescale $\tsf$ (as it includes also inflows and outflows), we can compare $t$ to $\tsf$ as an even more stringent requirement. In our model, $\tsf=\tdyn/\epsSFR$, which can be connected to the cosmic time if the disc radius is roughly a constant fraction $\lambda$ of the virial radius, via the virial dynamical time: $\tdyn\approx0.1\lambda t$. For $\lambda\approx0.05$, and our fiducial $\epsSFR=0.02$, we obtain $\tau<\tsf\approx0.25t$. Even if the star-formation timescale does not become shorter at higher redshift as we assume, and is instead roughly constant $\tsf\approx2\Gyr$ \citep{GenzelR_10a,DaddiE_10b}, $\tau$ is still short enough compared to the cosmic time during the epoch of interest here $z\leq2$, or even at somewhat higher redshifts.

In the discussion above, we have referred to the evolution of the average accretion rate over an ensemble of halos. In principle, the accretion rate of any individual galaxy may differ significantly from this average, thus preventing it from reaching the steady-state solution. However, both theoretical arguments and observational evidence indicate that this is likely not the case for the majority of galaxies. From a theoretical point of view, an upper limit on the variations of the accretion rates in individual galaxies is inferred from the scatter around the mean accretion rate found in cosmological simulations, which is $\approx0.2-0.3\rm{dex}$ for either dark matter \citep{GenelS_08a} or baryons \citep{DekelA_09a}. Indeed, galaxies in cosmological hydrodynamical simulations are often found to follow the steady-state solution for long cosmological times (e.g.~\citealp{CeverinoD_09a,GenelS_11a,DekelA_12a}). On the observational side, galaxies populate a rather narrow region in the SFR-stellar mass plane with an overall scatter of $\approx0.3\rm{dex}$ (e.g.~\citealp{DaddiE_07a,NoeskeK_07a,WhitakerK_12a,SalmiF_12a}). Such a tight relation could not be explained in the presence of large SFR variations.

We now consider the importance of mergers in driving galaxies out of the steady-state solution. It is clear that our model does not apply to galaxies after experiencing a major merger event, following which they go out of equilibrium in various ways, and in particular may have their disks transformed into spheroids. However, major mergers are not frequent, as minor mergers and smooth accretion contribute most of the mass in structure formation (e.g.~\citealp{GenelS_10a}). Minor mergers may trigger the formation of a bar, which is however a `slow' secular process that does not affect the processes we consider in this paper to a large degree, at least during the violent disk instability phase. They also do not produce significant star-formation bursts (e.g.~\citealp{JogeeS_09a}). Furthermore, cosmological simulations show that about half of the external galaxies that come in as minor mergers join the disk and share its kinematics without affecting its steady state \citep{MandelkerN_12a}. However, even if we consider all mergers with mass ratios $>1:10$ as `disruptive' as far as our model assumptions go, we should note that galaxies with a stellar mass of $10^{12}\Msun$ ($10^{11}\Msun$) at $z=0$ have probably experienced only $\approx7$ ($\approx1$) such mergers since $z=2$ \citep{HopkinsP_09a}. Considering this frequency against the number of galaxy dynamical times elapsed within this cosmic time ($>100$), we can expect most galaxies not to be found in an out-of-equilibrium state caused by mergers.

\section{Conversion of potential energy during disc inflows}
\label{s:inflows_efficiency_appendix}
The conversion of potential gravitational energy into turbulent energy ($\xi_m>0$) is justified by assuming the gas joins the outskirts of the disc and generates turbulence as it migrates to the bulge through the disc \citep{KrumholzM_10b,CacciatoM_11a}, rather than by hitting the disc from above. In this picture, the inflow rate $\gaminf$ adjusts itself to give $Q=1$ via a self-regulation loop, since the inflow is a result of the instability in the disc.

In principle, that released energy can be transferred in part to the dark matter component, but we do not address this explicitly here, as the dark matter is assumed to be subdominant inside the disc. The `immediate' radiation of this energy is not considered, because such losses are already encapsulated in the dissipation timescale of the turbulence. However, when the energy is transferred to the gas mass that stays in the disc, it can also in principle be in the form of rotational energy. The purpose of the following analysis is to show that under the standard picture of angular momentum loss, where a small mass gains most of the lost angular momentum, the conversion to rotational energy is negligible.

The angular momentum loss rate is $\dot{M}_{\rm inf}RV$. Let us assume that angular momentum is transferred into a fraction $a$ of the remainder of the disc mass. That fraction of mass that is pushed out by obtaining additional angular momentum has radius $R_a$ and velocity $V_a$ that can be different from that of the disc as a whole. This can be represented by writing
\bea
\label{e:angular_momentum}
\dot{M}_{\rm inf}RV &=& aM_{\rm g}\frac{\partial(R_aV_a)}{\partial t} = aM_{\rm g}\dot{R}_aV_a+aM_{\rm g}R_a\dot{V}_a \\\nonumber
&=& aM_{\rm g}\dot{R}_aV_a+aM_{\rm g} R_aV_a\frac{\dot{R}_a}{2R_a} = \frac{3}{2}aM_{\rm g}\dot{R}_aV_a,
\eea
where the penultimate equality is based on the equality $\dot{V}_a=0.5\sqrt{GM_{\rm g}}\dot{R}_a/(R_a)^{1.5}=0.5V_a\dot{R}_a/R_a$, which stems from the assumption $V_a=\sqrt{GM_{\rm g}/R_a}$. When the mass $aM_{\rm g}$ gains the angular momentum, it is accompanied by an energy gain rate that can be written as follows:
\bea
\label{e:energy_gain_from_angular_momentum}
\frac{\partial(GaM_{\rm g}^2/R_a)}{\partial t} &=& \frac{GaM_{\rm g}^2}{R_a^2}\dot{R}_a\\\nonumber
&=&aM_{\rm g}\frac{\dot{R}_a}{R}\frac{GM_{\rm g}}{R}(\frac{R}{R_a})^2\\\nonumber
&=&\frac{2}{3}\frac{V}{V_a}\dot{M}_{\rm inf}V^2(\frac{R}{R_a})^2.
\eea
This quantity should be compared with the energy that is available from the inflows, i.e.~$\dot{M}_{\rm inf}V^2$. The crucial factor is $a$, which comes in to the final expression in \equ{energy_gain_from_angular_momentum} via $(R/R_a)^2$. If the angular momentum is given to a small amount of mass ($a\ll1$) that is expelled to large distances due to the angular momentum it acquires, namely $R\ll R_a$, then the contribution of the inflows to rotational energy in the remainder of the disc is small, and most of the energy goes to turbulent energy ($\xi_m\approx1$). However, if the angular momentum of the inflowing gas is distributed evenly across the whole disc, then the rotational energy gain that is associated with this angular momentum gain is comparable to the energy that is released by the inflow, namely little energy is left for driving turbulence in the disc and then $\xi_m\ll1$.

\section{Approximations for parameter numerical values}
\label{s:param_values}
The following assumptions are used throughout in order to evaluate
numerical values for different quantities.
\begin{itemize}
\item
The star-formation efficiency per dynamical time is small: $\epsSFR\ll1$.
\item
The wind mass-loading factor is not very large: $\epsSFR^{-1}\gg(1+\etaw)$.
\item
Turbulent energy dissipation is much faster than star formation: $\tdis\approx\tdyn$.
\item
Angular momentum loss is {\it not} much faster than star formation: $\gaminf\sim\epsSFR$. Also, one can derive $\gaminf\approx0.01$ by comparing eqs.~(19)-(21) in \citet{DekelA_09b} to the definition of $\dot{M}_{\rm inf}$ in Section \ref{s:mass_steady_state}.
\item
When gas is accreted at the virial radius, it has potential energy of $\approx3V_{\rm vir}^2$ relative to the halo centre (for NFW haloes), and kinetic energy per unit mass of $\approx0.5V_{\rm in}^2$. We assume that most of that potential energy is lost to radiation (supported by simulations; \citealp{DekelA_09a}). Thus, at the arrival to the disc, the potential energy between the disc edge and the centre is $\approx V_{\rm vir}^2$ and the kinetic energy is still $\approx0.5V_{\rm in}^2$. \citet{DekelA_09a} estimate $V_{\rm in}\approx(1.5-2)V_{\rm vir}$. We assume in addition $\Vrot\approx(1-1.5)V_{\rm vir}$. Thus, $u\equiv\Vrot/V_{\rm in}\approx0.75\pm0.25$. We note that we thus assume roughly equal amounts of energy of the incoming gas, at the arrival to the disc, in kinetic and potential form: potential energy of $\approx V_{\rm vir}^2$ and kinetic energy of $\approx0.5V_{\rm in}^2=0.5(1.5-2)^2V_{\rm vir}^2\approx(1-2)V_{\rm vir}^2$.
\end{itemize}

\label{lastpage}

\end{document}